\documentstyle[epsf,multicol,prl,aps,eqsecnum]{revtex}
\newcommand{\calH}{{\cal H}}
\newcommand{\calD}{{\cal D}}
\newcommand{\calL}{{\cal L}}
\newcommand{\calZ}{{\cal Z}}
\newcommand{\calO}{g}
\newcommand{\calo}{o}
\newcommand{\tr}{{\rm tr\,}}
\newcommand{\Tr}{{\rm Tr\,}}
\newcommand{\Ln}{{\rm Ln\,}}
\newcommand{\trns}{\,^{\rm t}}
\newcommand{\sqrts}[1]{s^{#1\frac{1}{2}}}
\newcommand{\Qm}{Q_{\rm L}}
\newcommand{\Pm}{P_{\rm L}}
\newcommand{\Rm}{L}
\newcommand{\cc}{g_0}
\newcommand{\Ad}{A_{{\rm d}}^{-1}}
\newcommand{\Aod}[1]{A_{{\rm od}\,#1}^{-1}}
\newcommand{\intx}[1]{\int\!\!d^2#1}
\newcommand{\intp}[1]{\int\!\!\frac{d^2#1}{(2\pi)^2}}

\newcommand{\diff}[2]{\frac{{\rm d}#1}{{\rm d}#2}}
\newcommand{\ak}{{\tilde{a}}}

\begin{document}
\draft
\preprint{}
\title{Critical behavior of 
two-dimensional random hopping fermions with $\pi$-flux} 
\author{ Takahiro Fukui\cite{Email}} 
\address{Institut f\"ur Theoretische Physik,
Universit\"at zu K\"oln, Z\"ulpicher Strasse 77,
50937 K\"oln, Germany}
\date{February 26, 1999}
\maketitle
%----------------------------------------------------------------------
%                              Abstract
%----------------------------------------------------------------------
\begin{abstract}
A two dimensional random hopping model 
with $N$-species and $\pi$-flux is studied.
The field theory at the band center is shown to be in the 
universality class of GL($4m$,R)/O($4m$) nonlinear sigma model.
Vanishing beta function suggests delocalised states at 
the band center.
Contrary to the similar universality class with broken time reversal
symmetry, the present class is expected to have
at least two fixed point.
Large $N$-systems are shown to be in the weak-coupling fixed point,
which is characterized by divergent density of state, 
while small $N$ systems may be in the strong-coupling fixed point.
\end{abstract}
%\pacs{PACS: 72.15.Rn, 05.10.Cc, 11.10.-z\\
%{\it Keywords}: Anderson localization; Random hopping;
%Particle-hole symmetry; 
%Nonlinear sigma model; Renormalization group}
%\vspace{10mm}

\begin{multicols}{2}%---------------------------------------------!!!
%---------------------------------------------------------------------
%                      I. INTRODUCTION
%---------------------------------------------------------------------
\section{Introduction} \label{s:Int}

It is widely accepted that the metallic states are 
unstable in two dimensional (2D) disordered systems\cite{AALR}.
However, two typical exceptions have been known
for a couple of decades.
One is the system with strong spin-orbit coupling \cite{HLN} and
the other is the integer quantum Hall (IQH) 
system \cite{And,Pru,Huc}.
%which are described, respectively, 
%by the symplectic and by the unitary
%nonlinear sigma model (NLSM).
Especially, IQH transition has attracted much interest, because,
as shown by Pruisken {\it et al.}\cite{Pru},
what is responsible for the delocalization is a 
topological term in the usual nonlinear sigma model 
(NLSM) description.
Though his arguments clarified nonperturbative aspects of
the delocalization phenomena in the IQH systems,
it was not generally possible to calculate critical properties of 
the IQH transitions.
In order to calculate them explicitly, a model of 
Dirac fermions with various types of disorder
was studied extensively\cite{LFSG,Ber,MCW,KMT,HoCha}. 
This model has also intimate relationship with
a statistical model \cite{DotDot,Lud,Sha}
and with more physical systems such as
$t$-$J$ model \cite{AffMar} or 
$d$-wave superconductor\cite{Lee,BFN,SFBN}.
However, many problems remain to be explored, 
since generic IQH transition is in the strong-coupling 
limit\cite{LFSG}, 
though some exact results can be obtained for models with
rather limited realization of disorder, e.g. the model only with
random vector potentials\cite{LFSG}.

On the other hand, Gade found a new universality class 
having a random critical point\cite{Gad}. 
It is due to what is called particle-hole symmetry,
which enlarges the symmetry of the universality class to
the general linear group.
The renormalization group equations were
calculated for the NLSM specifying such universality class 
with \cite{GadWeg} or without 
\cite{Gad,GadWeg} time-reversal symmetry.  
Recently, it was recognized that the random flux model 
belongs to this universality class\cite{Fur} or to its
supersymmetric version\cite{AltSim}.

Recently, Hatsugai {\it et al.} (HWKM)\cite{Hat} proposed 
an interesting 2D model with the particle-hole symmetry.
This model describes {\it random hopping} fermions 
on the square-lattice with $\pi$-flux.
It should have close relationship with 
various kinds of systems mentioned above, since
it is described by the Dirac fermions near the band center.
Numerical calculations by HWKM strongly suggest a random critical
point at the band center.
This model is actually in sharp contrast to the similar model
but with on-site disorder\cite{FisFra} which 
has only localized states.
Especially, what is interesting is the power-law behavior of 
the density of state (DOS) such as $\rho\sim E^\nu$ with $\nu$
dependent on the strength of the disorder.
Their calculations show that the exponent $\nu$ keeps
positive even for rather large randomness.
This behavior of the DOS seems
to be in contrast to the pioneering work 
on the two-sublattice model by
Oppermann and Wegner\cite{OppWeg}, 
where the two-particle Green function
has singularity at the band center.
Even in one dimension, 
this symmetry plays a role in the existence 
of the critical point\cite{TheCoh,EggRie,DSFis,Mck,BalFis},
which brings about a singularity of the DOS at the band center.
%Balents and Fisher \cite{BalFis} obtained some
%exact results for the random hopping model
%to clarify the delocalization observed so far\cite{DSFis,Mck}, 
%using supersymmetric quantum mechanics.
Therefore, it is interesting to study the 2D model proposed
by HWKM, especially paying attention to the 
behavior of DOS.

In this paper, we study the 2D random hopping model with $N$-species.
The model with one species corresponds to the HWKM model.
In order to take $\pi$-flux into account effectively,
we firstly derive the Dirac Hamiltonian
and next apply the NLSM method developed by Gade\cite{Gad}.
We show that the model can be described by the NLSM on 
the symmetric space GL($4m$,R)/O($4m$),
which is just the class {\it BD}I by Zirnbauer\cite{Zir}. 
We conclude that the band center is a random critical point.
However, contrary to the class with broken time-reversal symmetry,
the present model should have, at least, two fixed point,
which are characterized by the scaling property of the DOS.
It is shown that large $N$ systems are in the weak-coupling
fixed point, where the DOS diverges. A discontinuity 
is, therefore, expected at zero disorder.
It is also conjectured that the $N=1$ model proposed by HWKM, 
or more generally, small $N$ systems
are in the strong-coupling fixed point.

In Section \ref{s:Mod}, we introduce the model and derive its 
continuum limit. In Section \ref{s:Rep}, we calculate the 
generating functional of the Green functions and discuss the 
symmetry of the model. In Section \ref{s:Aux}, we use the trick of 
auxiliary field and give the saddle point solution.
Section \ref{s:Der} is devoted to the derivation of
the nonlinear sigma model describing the Goldstone mode.
Section \ref{s:Ren} deals with the renormalization group.
Summary is given in Section \ref{s:Sum}.

It should be noted that 
the following calculations are indebted to
\cite{WegSch,PruSch} and especially to\cite{Gad}. 

%---------------------------------------------------------------------
%                             MODEL
%---------------------------------------------------------------------
\section{Model} \label{s:Mod}

We present a slightly generalized model of HWKM,
introducing an additional internal degrees of freedom.
The tight-binding Hamiltonian is defined by
%----------------------------------------------------------------------
%   lattice Hamiltonian
%----------------------------------------------------------------------
\begin{equation}
H=\sum_{\langle i,j\rangle}\sum_{\sigma,\tau}c_{i\sigma}^\dagger 
t_{i\sigma,j\tau}c_{j\tau} +\mbox{h.c.},
\label{TigHam}%--------------------------------------------------------
\end{equation}
where $j=(j_x,j_y)$ and $\sigma=1,2,\cdots,N$ denote, respectively,
a site on the square lattice and a species of fermions.
The summation with respect to sites is over
nearest-neighbor pairs.
The hopping matrix elements are defined by 
%----------------------------------------------------------------------
%   hopping elements
%----------------------------------------------------------------------
\begin{eqnarray}
&& 
t_{j+\hat{x}\sigma,j\tau}=
(-)^{j_y}t\delta_{\sigma\tau}+\delta t_{x,\sigma\tau},\\
&&
t_{j+\hat{y}\sigma,j\tau}=
t\delta_{\sigma\tau}+\delta t_{y,\sigma\tau},
\end{eqnarray}
with $\hat{x}=(1,0)$ and $\hat{y}=(0,1)$.
Here, $\delta t_{x,\sigma\tau}$ and $\delta t_{y,\sigma\tau}$
are assumed to be independent random variables.
The $N=1$ model is just the one proposed by HWKM.
As stressed in \cite{Hat}, the point 
is that the model has randomness in the hopping terms, which
causes the particle-hole symmetry,
%--------------------------------------------------------------------
%   particle-hole symmetry
%--------------------------------------------------------------------
\begin{equation}
H\rightarrow -H \qquad
\mbox{for}\quad c_{j\sigma}
\rightarrow (-)^{j_x+j_y}c_{j\sigma}.
\label{ParHol}%----------------------------------------------------
\end{equation}
Accordingly, if there is an eigenstate of energy $e$, 
there also exists a conjugate pair with energy $-e$.
The zero energy states is, however, special, because
there is no partner for it.

Let us first consider the pure model without randomness.
In this case, particles with different species do not couple 
each other. 
In the momentum space, the Hamiltonian for each species is
\end{multicols}\noindent%------------------------------------------!!!
%----------------------------------------------------------------------
%   Hamiltonian in momentum space
%----------------------------------------------------------------------
\begin{eqnarray}
H_\sigma&&=2t
\sum_{\stackrel{\scriptstyle{-\pi\le ak_x<\pi}}{0\le ak_y<\pi}}
\left(c_{\sigma k}^\dagger,c_{\sigma k+k_0}^\dagger\right)
\left(
 \begin{array}{rr}
  \cos(ak_y)&\cos(ak_x)\\ \cos(ak_x)&-\cos(ak_y)
 \end{array}
\right)
\left(\begin{array}{c}c_{\sigma k}\\ c_{\sigma k+k_0}\end{array}\right),
\end{eqnarray}
where $a$ is the lattice constant and $k_0=(0,\pi/a)$.
Eigenvalues are $E=\pm 2t\sqrt{\cos^2ak_x+\cos^2ak_y}$, 
and hence there exist two Fermi points given by
$k_F=(\pm\pi/2a,\pi/2a)$.
In order to investigate the model near the band center,
it is convenient to linearize the dispersion at the two Fermi points
and take a continuum limit\cite{FisFra,Hat}.
This may be valid up to a certain energy scale specified 
by a cut-off $\Lambda$. 
The lattice operator takes the following form in the continuum limit, 
%----------------------------------------------------------------------
%   continuum Fermi operator
%----------------------------------------------------------------------
\begin{eqnarray}
c_{j\sigma}/a\sim\psi_\sigma(x)= &&
 i^{j_x+j_y}\psi_{\sigma1}(x)
+i^{j_x-j_y}\psi_{\sigma2}(x)
%\nonumber\\&&
+i^{-j_x+j_y}\psi_{\sigma3}(x)
+i^{-j_x-j_y}\psi_{\sigma4}(x),
\end{eqnarray}
where $x=aj$.

\begin{multicols}{2}%----------------------------------------------!!!
Now it is easy to derive the total continuum Hamiltonian including 
effects of the random hopping. The result is
%----------------------------------------------------------------------
%   continuum total Hamiltonian 
%----------------------------------------------------------------------
\begin{equation}
H=\intx{x}\Psi^\dagger\calH\Psi,
\end{equation} 
where 
$\Psi^\dagger=
(\Psi_1^\dagger,\cdots,\Psi_\sigma^\dagger,\cdots,\Psi_N^\dagger)$ 
with
$
\Psi_\sigma^\dagger=
(
\psi_{\sigma1}^\dagger,
\psi_{\sigma2}^\dagger,
\psi_{\sigma3}^\dagger,
\psi_{\sigma4}^\dagger
)
$,
and
%---------------------------------------------------------------------
%   continuum one-particle Hamiltonian
%---------------------------------------------------------------------
\begin{eqnarray}
&&
\calH=\calH_{\rm p}+\calH_{\rm d},
\nonumber\\
&&
\left(
\calH_{\rm p}
\right)_{\sigma\sigma'}=vh_{\rm p}\delta_{\sigma\sigma'};
 \quad
h_{\rm p}=\alpha_\mu i\partial_\mu,\quad v=2at,
\nonumber\\
&&
\left(
\calH_{\rm d}
\right)_{\sigma\sigma'}
\nonumber\\&&\quad%-----------------------------------------------!!!
=\frac{1}{2}\sum_{j=1}^4
\left[
\left(
 V^j_{\sigma\sigma'}+V^j_{\sigma'\sigma}
\right)\gamma_j
+
\left(
 U^j_{\sigma\sigma'}-U^j_{\sigma'\sigma}
\right)i\alpha_j
\right]
\end{eqnarray}
Here, $\calH_{\rm p}$ and $\calH_{\rm d}$
describe the pure and the disorder Hamiltonian, respectively, 
and $\mu=1,2$.
$V^j_{\sigma\sigma'}$ and $U^j_{\sigma\sigma'}$ ($j=1,\ldots,4$) 
are independent random variables associated with the 
Fourier components of 
$\delta t_{x,\sigma\sigma'}$ and $\delta t_{y,\sigma\sigma'}$, 
assumed to obey the Gaussian distribution,
%----------------------------------------------------------------------
%   distribution of randomness
%----------------------------------------------------------------------
\begin{eqnarray}
P[V^j_{\sigma\sigma'}]\propto
\int\calD[V]e^{-\frac{N}{\cc}\intx{x}(V^j_{\sigma\sigma'})^2},
\nonumber\\
P[U^j_{\sigma\sigma'}]\propto
\int\calD[U]e^{-\frac{N}{\cc}\intx{x}(U^j_{\sigma\sigma'})^2}.
\label{RanDis}%--------------------------------------------------------
\end{eqnarray}
Matrices $\alpha$'s and $\gamma$'s are defined by
%----------------------------------------------------------------------
%   \alpha and \gamma matrices in standard basis
%----------------------------------------------------------------------
\begin{eqnarray}
&&
\alpha_1=\sigma_1\otimes\tau_3, \quad
\alpha_2=\sigma_3\otimes1 ,     \quad
\nonumber\\
&&
\alpha_3=1       \otimes\tau_3 ,\quad
\alpha_4=\sigma_3\otimes\tau_1, \quad
\nonumber\\
&&
\gamma_1=1\otimes\tau_2,       \quad
\gamma_2=\sigma_2\otimes\tau_1,\quad
\nonumber\\
&&
\gamma_3=\sigma_1\otimes\tau_2,\quad
\gamma_4=\sigma_2\otimes1,     \quad
\nonumber\\
&&
\gamma=\sigma_1\otimes\tau_1.
\label{StaBas}%--------------------------------------------------------
\end{eqnarray}
The particle-hole transformation Eq. (\ref{ParHol}) is 
expressed by $\gamma$ as $\gamma H\gamma=-H$.
In the case of the
on-site disorder\cite{FisFra}, disorder potentials in the 
continuum limit 
commute each other and hence they are trivially diagonalized,
while in the random hopping case, four $\gamma$'s are noncommutative. 
Moreover, there appear extra disorder terms described by matrices
$\alpha$'s caused by multi-species.
When $N=1$, $\alpha$-terms disappear and
this Hamiltonian reduces to the one derived by HWKM. 

In order to apply the same techniques as Gade's
to the present model,
it may be convenient to switch into the chiral basis, where
$\gamma$ is diagonal.
To this end, make a transformation, 
$\gamma\rightarrow U^\dagger\gamma U$, where
%----------------------------------------------------------------------
%   transformation matrix
%----------------------------------------------------------------------
\begin{equation}
U=
\frac{1}{2\sqrt{2}}
\left(
\begin{array}{rrrr}
  1+i& 1+i&1-i& 1-i\\
  1+i&-1-i&1-i&-1+i\\
 -1-i& 1+i&1-i&-1+i\\
 -1-i&-1-i&1-i& 1-i
\end{array}
\right).
\end{equation}
The matrices defined in Eq. (\ref{StaBas}) are then converted to
%----------------------------------------------------------------------
%   alpha and gamma matrices in chiral basis
%----------------------------------------------------------------------
\begin{eqnarray}
&&
\alpha_1=\sigma_3\otimes\tau_2,\quad
\alpha_2=\sigma_1\otimes\tau_2,\quad
\nonumber\\
&&
\alpha_3=1\otimes\tau_2,\quad
\alpha_4=-\sigma_2\otimes\tau_1,\quad
\nonumber\\
&&
\gamma_1=-\sigma_3\otimes\tau_1,\quad
\gamma_2=-\sigma_1\otimes\tau_1,\quad
\nonumber\\
&&
\gamma_3=-1\otimes\tau_1,       \quad
\gamma_4=-\sigma_2\otimes\tau_2,\quad
\nonumber\\
&&
\gamma=-1\otimes\tau_3.
\label{ChiBas}%-------------------------------------------------------
\end{eqnarray}
In the new basis, the Hamiltonian is real and symmetric
due to the time-reversal symmetry.
Moreover, the Hamiltonian 
has now only off-diagonal elements in $\tau$-space, 
as is expected for systems with the particle-hole symmetry.
The $\tau$-space is thus associated with the $\pm$-sublattices 
in the lattice model (\ref{TigHam}).

%---------------------------------------------------------------------
%                       REPLICA METHODS
%---------------------------------------------------------------------
\section{Replica methods}\label{s:Rep}

In this section, we will 	firstly derive the connection of 
the Green functions between the lattice 
and the continuum theory\cite{BalFis}, 
and secondly develop the continuum theory using the replica method.

%---------------------------------------------------------------------
%                       GREEN FUNCTION
%---------------------------------------------------------------------
\subsection{Green function} \label{s:Gre}

Single-particle (e.g. retarded) Green function of the lattice system
is given by 
%---------------------------------------------------------------------
%   lattice Green function
%---------------------------------------------------------------------
\begin{equation}
G_L(j\sigma,j'\sigma';z)
=-i\int_0^\infty dte^{iz}
\langle0|c_{j\sigma}(t)c_{j'\sigma'}^\dagger(0)|0\rangle,
\end{equation}
where $z=E+i\omega$, $c_{j\sigma}(t)=e^{iHt}c_{j\sigma}e^{-iHt}$
and $|0\rangle$ denotes the fermion vacuum. 
This Green function is expressed in terms of the continuum fields as
%---------------------------------------------------------------------
%   relation between lattice and continuum Green function
%---------------------------------------------------------------------
\begin{eqnarray}
&&G_L(j\sigma,j'\sigma';z)
\sim e^{i\frac{\pi}{a}(x-x'+y-y')}
\nonumber\\&&\qquad\times%----------------------------------------!!!
\sum_{\alpha=1}^4 
e^{i\frac{\pi}{a}[n_\alpha(x-x')+m_\alpha(y-y')]}
G_{\alpha\alpha}(x\sigma,x'\sigma';z) ,
\label{RelGre}%-------------------------------------------------------
\end{eqnarray}
where constant vectors in oscillating phases factors are
${\bf n}=(0,1,0,1)$ and ${\bf m}=(0,0,1,1)$, and where
%---------------------------------------------------------------------
%   continuum Green function
%---------------------------------------------------------------------
\begin{eqnarray}
G_{\alpha\beta}(x\sigma,x'\sigma';z)
%\nonumber\\
%&&\qquad
&&=-i\int_0^\infty dte^{iz}
\langle0|
 \psi_{\sigma\alpha}(x,t)
\psi_{\sigma'\beta}^\dagger(x',0)
|0\rangle
\nonumber\\
&&=\langle x\sigma\alpha|(z-\calH)^{-1}|x'\sigma'\beta\rangle.
\end{eqnarray}
In Eq. (\ref{RelGre}), terms 
breaking translational invariance, i.e., off-diagonal 
Green functions $G_{\alpha\beta}$ with $\alpha\ne\beta$ are 
neglected.

Since we are mainly interested in the DOS for the lattice system, 
%----------------------------------------------------------------------
%   DOS for lattice system
%----------------------------------------------------------------------
\begin{eqnarray}
\rho(E)=\mp\frac{1}{\pi}\lim_{\omega\rightarrow+0}
\sum_\sigma{\rm Im\,} G_{\rm L}(j\sigma,j\sigma,E\pm i\omega),
\end{eqnarray}
let us define the following combinations of the Green functions
for the continuum theory,
%----------------------------------------------------------------------
%   Special combination of Green function for continuum theory
%----------------------------------------------------------------------
\begin{eqnarray}
&&
G^k(x)=\sum_{\sigma}\sum_{\alpha=1}^4
G_{\alpha\alpha}(x\sigma,x\sigma,z_k) ,
\nonumber\\
&&
K^{kk'}(x,x')=\sum_{\sigma,\sigma'}\sum_{\alpha,\alpha'=1}^4
G_{\alpha\alpha'}(x\sigma,x'\sigma',z_k) 
\nonumber\\&&\qquad\qquad\qquad\qquad\qquad\qquad\times%-----------!!!
G_{\alpha'\alpha}(x'\sigma',x\sigma,z_{k'}) ,
\label{ConGre}%--------------------------------------------------------
\end{eqnarray}
where $k$ specify the retarded and advanced Green function.
(See Eq. (\ref{Defzs}) for more precise definition.)
%----------------------------------------------------------------------
%                    Replica field theory
%----------------------------------------------------------------------
\subsection{Generating functional of the Green functions}
\label{s:RepFie}

Since the Hamiltonian is real and symmetric in the chiral basis,
we can apply the real bosonic replica method.
Generating functional of the Green functions is introduced as 
%----------------------------------------------------------------------
%   partition function
%----------------------------------------------------------------------
\begin{equation}
\calZ=\int\calD[\Phi]e^{-S-S_{\rm s}},
\end{equation}
where
%----------------------------------------------------------------------
%   action
%----------------------------------------------------------------------
\begin{equation}
S=\frac{1}{2}\intx{x}\sum_{a,k}s_k\trns\Phi_a^{k}(z_k-\calH)\Phi_a^k,
\end{equation}
with $k=(p,q)$,
%----------------------------------------------------------------------
%   z_k and s_k
%----------------------------------------------------------------------
\begin{eqnarray}
&&z_k=(-)^{q+1}E+(-)^{p+1}\omega
=(-)^{q+1}E-s_k\omega,
\nonumber\\
&&s_k=-i\mbox{sgn\,}{\rm Im\,} z_k=(-)^pi.
\label{Defzs}%---------------------------------------------------------
\end{eqnarray}
Indices $a=1,\ldots,m$ of scaler field $\Phi_a^{k}$ 
denotes the replica.
In what follows, to simplify the notations, we sometimes use
$\ak=(a,k)$ and $\ak'=(a',k')$, and hence, e.g., $\Phi_a^k=\Phi_\ak$.
Fixing $\ak$,  the field $\Phi_\ak$ is multi-component with
respect to the species
$
\trns\Phi_a^{k}\equiv
(\trns\Phi_{1\ak},\ldots,\trns\Phi_{\sigma\ak},\ldots,
\trns\Phi_{N\ak})
$
and to Dirac indices
$\trns\Phi_{\sigma\ak}=(\Phi_{\sigma\alpha\ak})$
with $\alpha=1,\ldots,4$.
$S_{\rm s}$ is a source term which will be introduced momentarily.
Green functions with quenched disorder is expressed by
%---------------------------------------------------------------------
%   Green function with quenched disorder
%---------------------------------------------------------------------
\begin{eqnarray}
&&
G^k(x)=\lim_{m\rightarrow0}
 s_k\sum_\sigma\sum_{\alpha=1}^4
\langle
 \Phi_{\sigma\alpha \ak}(x)
 \Phi_{\sigma\alpha \ak}(x)
\rangle ,
\nonumber\\
&&
K^{kk'}(x,x')=\lim_{m\rightarrow0}
 s_ks_{k'}\sum_{\sigma,\sigma'}\sum_{\alpha,\alpha'=1}^4
\nonumber\\&&\qquad\qquad%----------------------------------------!!!
\langle
 \Phi_{\sigma'\alpha'\ak }(x')
 \Phi_{\sigma \alpha \ak }(x)
 \Phi_{\sigma \alpha \ak'}(x)
 \Phi_{\sigma'\alpha'\ak'}(x')
\rangle,
\end{eqnarray}
where $\langle\cdots\rangle$ is 
the expectation value with respect to $S$.
Since the two components of the $\tau$-space in Eq. (\ref{ChiBas}) 
play an important role in Gade's argument\cite{Gad},
let us introduce two fields $\phi^+$ and $\phi^-$ explicitly,
each of which has Dirac indices 
$\alpha=1,2$ of the $\sigma$-space, i.e.,
$(\Phi_{\sigma1\ak},\Phi_{\sigma2\ak},
  \Phi_{\sigma3\ak},\Phi_{\sigma4\ak})
=(\phi_{\sigma1\ak}^+,\phi_{\sigma2\ak}^+,
  \phi_{\sigma1\ak}^-,\phi_{\sigma2\ak}^-,)$.
The two fields $\phi^\pm$ reflect the 
particle-hole symmetry in the original lattice Hamiltonian, 
and are referred to as $r=\pm$ fields.
To be more concrete, let us write down the Lagrangian
with respect to $\phi^\pm$ fields,
%----------------------------------------------------------------------
%   Lagrangian
%----------------------------------------------------------------------
\begin{eqnarray}
&&
\calL=\calL_{\rm p}+\calL_{\rm b}+\calL_{\rm d}+\calL_{\rm s},
\nonumber\\
&&
\calL_{\rm p}+\calL_{\rm b}=\frac{1}{2}\sum_{\ak}s_k\sum_{\sigma}
\nonumber\\&&\qquad%----------------------------------------------!!!
\left(
 \trns\phi_{\sigma \ak}^+,\trns\phi_{\sigma \ak}^-
\right)
\left(
 \begin{array}{rr}z_k&-h_0\\h_0&z_k\end{array}
\right)
\left(
 \begin{array}{c}
  \phi_{\sigma \ak}^+\\
  \phi_{\sigma \ak}^-
 \end{array}
\right),
\label{LagPB}\\%-------------------------------------------------------
&&
\calL_{\rm d}=
-\frac{1}{2}\sum_{\ak}s_k\sum_{\sigma,\sigma'}\sum_{j=1}^4
\nonumber\\&&\quad%-----------------------------------------------!!!
\Bigg[
U_{\sigma\sigma'}^j
\left(
 \trns\phi_{\sigma \ak}^+,\trns\phi_{\sigma \ak}^-
\right)
\left(
 \begin{array}{cc}
  0&\widetilde\gamma_j\\
  -\trns\widetilde\gamma_j&0
 \end{array}
\right)
\left(
 \begin{array}{c}
  \phi_{\sigma' \ak}^+\\\phi_{\sigma' \ak}^-
 \end{array}
\right)
\nonumber\\&&\quad%------------------------------------------------!!!
-V_{\sigma\sigma'}^j
\left(
 \trns\phi_{\sigma \ak}^+,\trns\phi_{\sigma \ak}^-
\right)
\left(
 \begin{array}{cc}
  0&\widetilde\gamma_j\\
  \trns\widetilde\gamma_j&0
 \end{array}
\right)
\left(
 \begin{array}{c}
  \phi_{\sigma' \ak}^+\\\phi_{\sigma' \ak}^-
 \end{array}
\right)
\Bigg],
\label{LagD}\\%--------------------------------------------------------
&&
\calL_{\rm s}=
-\sum_{\ak,\ak'}s_k^{\frac{1}{2}}s_{k'}^{\frac{1}{2}}
\sum_{\sigma}
\nonumber\\&&\qquad%-----------------------------------------------!!!
\left(
 \trns\phi_{\sigma \ak}^+,\trns\phi_{\sigma \ak}^-
\right)
\left(
 \begin{array}{cc}h_{\ak\ak'}^+&0\\0&h_{\ak\ak'}^-\end{array}
\right)
\left(
 \begin{array}{c}
  \phi_{\sigma \ak'}^+\\\phi_{\sigma \ak'}^-
 \end{array}
\right),
\label{LagS}%----------------------------------------------------------
\end{eqnarray}
where
%----------------------------------------------------------------------
%   h_0 and \widetilde\gamma
%----------------------------------------------------------------------
\begin{eqnarray}
&&
h_0=\widetilde\gamma_\mu\partial_\mu,
\nonumber\\
&&
\widetilde\gamma_1=\sigma_3,  \quad
\widetilde\gamma_2=\sigma_1,  \quad
\widetilde\gamma_3=1,         \quad
\widetilde\gamma_4=-i\sigma_2,\quad
\end{eqnarray}
and $\calL_{\rm p}$, $\calL_{\rm b}$, 
$\calL_{\rm d}$, and $\calL_{\rm s}$
describe
pure-, breaking-, disorder-, and source-term, 
respectively.
We set $v=1$ without loss of generality.
The parameters in the source terms are symmetric
$h_{\ak\ak'}(x)=h_{\ak'\ak}(x)$
and are, more explicitly, given by
$h^\pm\,\!_{aa'}^{kk'}$.
 
Let us also divide the Green functions in Eq. (\ref{ConGre})
into two kinds of contributions from $\phi^\pm$,
and average them over disorder (\ref{RanDis})
%----------------------------------------------------------------------
%   Green functions
%----------------------------------------------------------------------
\begin{eqnarray}
&&
\overline{G}^k(x)=\sum_{r=\pm}\overline{G}_r^k(x),
\nonumber\\
&&
\overline{K}^{kk'}(x,x')=\sum_{r,r'=\pm}\overline{K}_{rr'}^{kk'}(x,x'),
\end{eqnarray}
where
\begin{eqnarray}
\overline{G}_\pm^k(x)&&=\lim_{m\rightarrow0}
 s_k\sum_\sigma\sum_{\alpha=1}^2
\overline{
\langle
 \phi_{\sigma\alpha \ak}^\pm(x)
 \phi_{\sigma\alpha \ak}^\pm(x)
\rangle
         }
\nonumber\\
&&
=\lim_{m\rightarrow0}
\left.
\frac{\partial\calZ}{\partial h_{\ak\ak}^\pm(x)}
\right|_{h^+=h^-=0},
\nonumber\\
\overline{K}_{rr'}^{kk'}(x,x')&&=\lim_{m\rightarrow0}
 s_ks_{k'}\sum_{\sigma,\sigma'}\sum_{\alpha,\alpha'=1}^2
\nonumber\\&&%----------------------------------------------------!!!
\overline{
\langle
 \phi_{\sigma'\alpha'\ak }^{r }(x')
 \phi_{\sigma \alpha \ak }^{r'}(x)
 \phi_{\sigma \alpha \ak'}^{r'}(x)
 \phi_{\sigma'\alpha'\ak'}^{r }(x')
\rangle
         }
\nonumber\\
&&
=\lim_{m\rightarrow0}\frac{1}{4}
\left.
 \frac{\partial^2\calZ}
      {\partial h_{\ak\ak'}^{r'}(x)\partial h_{\ak'\ak}^{r}(x')} 
\right|_{h^+=h^-=0}.
\label{AveGre}%--------------------------------------------------------
\end{eqnarray}
Overbars denote ensemble-average, defined by
%----------------------------------------------------------------------
%   average of expectation value
%----------------------------------------------------------------------
\begin{eqnarray}
\overline{\langle O\rangle}=&&\lim_{m\rightarrow0}
\int\calD[V]\calD[U]\calD[\Phi]O
\nonumber\\&&\times%-----------------------------------------------!!!
e^{-S-\frac{N}{\cc}\intx{x}
  \sum_{\sigma,\sigma'}\sum_j
  \left[
  (V_{\sigma\sigma'}^j)^2+(U_{\sigma\sigma'}^j)^2
  \right]
  },
\label{BarExp}%--------------------------------------------------------
\end{eqnarray}
and hence, $\calZ$ in Eq. (\ref{AveGre}) should
be considered as including the disorder 
term in Eq. (\ref{BarExp}).
Integration over $V$ converts $\calL_{\rm d}$ 
into the form of four-point interactions,
%----------------------------------------------------------------------
%   4 point interaction
%----------------------------------------------------------------------
\begin{equation}
\calL_{\rm d}=
-\frac{\cc}{2N}\sum_{\ak,\ak'}s_ks_{k'}\sum_{\sigma,\sigma'}
\sum_{\alpha,\alpha'=1}^2
\phi_{\sigma \alpha  \ak }^+
\phi_{\sigma'\alpha' \ak }^-
\phi_{\sigma \alpha  \ak'}^+
\phi_{\sigma'\alpha' \ak'}^-.
\label{IntLag}%--------------------------------------------------------
\end{equation}

As Gade discussed\cite{Gad}, this Lagrangian has unique 
symmetry property, which is caused by the particle-hole
symmetry of the original model.
Namely, it is easily verified from 
Eqs. (\ref{LagPB}), (\ref{LagS}) and (\ref{IntLag}) that 
when $z_k=0$, the Lagrangian is invariant under the transformation
%----------------------------------------------------------------------
%   GL transformation
%----------------------------------------------------------------------
\begin{equation}
\phi_{\sigma\alpha}^+\rightarrow
 \calO\phi_{\sigma\alpha}^+,
\quad
\phi_{\sigma\alpha}^-\rightarrow 
 -s\trns\calO^{-1}s\phi_{\sigma\alpha}^-,
\label{GlTra}%---------------------------------------------------------
\end{equation}
where $\calO=\calO_{aa'}^{kk'}\in{\rm GL}(4m,{\rm R})$,
and $s=s_{aa'}^{kk'}$ is a matrix defined by
%----------------------------------------------------------------------
%   matrix s
%----------------------------------------------------------------------
$
s=\delta_{aa'}s_k\delta_{kk'}
$.
Since the symmetry-breaking terms are written as
%----------------------------------------------------------------------
%   Lagrangian for the discussion of symmetry
%----------------------------------------------------------------------
\begin{eqnarray}
&&
\calL_{\rm b}=\calL_\omega+\calL_E,
\nonumber\\
&&
\calL_\omega=\frac{\omega}{2}\sum_\sigma\sum_{\alpha=1}^2
\left(
  \trns\phi_{\sigma\alpha}^+\phi_{\sigma\alpha}^+
 +\trns\phi_{\sigma\alpha}^-\phi_{\sigma\alpha}^-
\right) ,
\nonumber\\
&&
\calL_E=-\frac{iE}{2}\sum_\sigma\sum_{\alpha=1}^2
\left(
  \trns\phi_{\sigma\alpha}^+\lambda\phi_{\sigma\alpha}^+
 +\trns\phi_{\sigma\alpha}^-\lambda\phi_{\sigma\alpha}^-
\right),
\label{SymLag}%--------------------------------------------------------
\end{eqnarray}
with 
$\lambda=\delta_{aa'}(-)^{p+q}\delta_{kk'}$,
GL($4m$,R) is broken by $\calL_\omega$ up to O($4m$).
For non-zero energy, $\calL_E$ implies that the symmetry
group is O($2m,2m$), which is broken to O($2m)\times$O($2m$)
by the $\calL_\omega$.
This suggests that all states with nonzero energies
are localized.

%--------------------------------------------------------------------
%                 AUXILIARY MATRIX FIELD THEORY  
%--------------------------------------------------------------------
\section{Auxiliary matrix field theory}\label{s:Aux}

\subsection{Auxiliary fields} \label{s:AuxFie}
In order to apply the well-known trick of auxiliary
fields, let us first introduce two kinds of fields,
%--------------------------------------------------------------------
%   \rho and \sigma
%--------------------------------------------------------------------
\begin{eqnarray}
&&
\rho_{aa'}^{kk'}=\frac{1}{2}\sum_\sigma\sum_{\alpha=1}^2
\left(
 \phi_{\sigma\alpha \ak}^+\phi_{\sigma\alpha \ak'}^+
+\phi_{\sigma\alpha \ak}^-\phi_{\sigma\alpha \ak'}^-
\right),
\nonumber\\
&&
\sigma_{aa'}^{kk'}=\frac{i}{2}\sum_\sigma\sum_{\alpha=1}^2
\left(
 \phi_{\sigma\alpha \ak}^+\phi_{\sigma\alpha \ak'}^+
-\phi_{\sigma\alpha \ak}^-\phi_{\sigma\alpha \ak'}^-
\right) .
\end{eqnarray}
By the use of these fields, we can express
%--------------------------------------------------------------------
%   
%--------------------------------------------------------------------
\begin{eqnarray}
&&
\calL_\omega=-\omega\,\tr s\sqrts{}\rho\sqrts{} ,
\nonumber\\
&&
\calL_{\rm s}=-\tr (h\sqrts{}\rho\sqrts{}
 -i\bar{h}\sqrts{}\sigma\sqrts{}) ,
\nonumber\\
&&
\calL_{\rm d}=-\frac{\cc}{2N}\tr
\left[
 (\sqrts{}\rho\sqrts{})^2+(\sqrts{}\sigma\sqrts{})^2
\right] ,
\end{eqnarray}
where 
$h=h^++h^-,\quad \bar{h}=h^+-h^-$, and
$\tr$ means the trace in terms of $a$ and $k$ indices, i.e.,
$\tr O=\sum_{a,k}O_{aa}^{kk}$.
Let us further define 
%--------------------------------------------------------------------
%   X and Y
%--------------------------------------------------------------------
\begin{eqnarray}
&&
X=\frac{\cc}{N}\sqrts{}\rho\sqrts{}+h +\omega s,
\nonumber\\
&&
Y=\frac{\cc}{N}\sqrts{}\sigma\sqrts{}-i\bar{h},
\end{eqnarray}
where $X$ and $Y$ are matrices with indices
$X_{aa'}^{kk'}$ and $Y_{aa'}^{kk'}$.
We are then ready to introduce auxiliary fields.
Noting the following identities,
%--------------------------------------------------------------------
%   identities
%--------------------------------------------------------------------
\begin{eqnarray}
&&
\calL_\omega+\calL_{\rm s}+\calL_{\rm d}
 +\frac{N}{2\cc}\tr\left(X^2+Y^2\right)
\nonumber\\&&\qquad%--------------------------------------------!!!
=\frac{N}{2\cc}\tr
\left(
 h^2-\bar{h}^2-\omega^2+2\omega hs
\right),
\nonumber\\
&&
\int\calD[Q]\calD[P]
e^{-\frac{N}{2\cc}\intx{x}\tr
\left[
 (Q-X)^2+(P-Y)^2
 \right]
}=1,
\end{eqnarray}
\end{multicols}\noindent%----------------------------------------!!!
where $Q$ and $P$ are real and symmetric matrices,
we have the following expression for the generating functional
depending on auxiliary fields $Q$ and $P$ as well as
$\phi^\pm$,
%--------------------------------------------------------------------
%   tilde Lagrangian
%--------------------------------------------------------------------
\begin{eqnarray}
\calZ=&&\int\calD[\phi^+]\calD[\phi^-]\calD[Q]\calD[P]
 e^{-\intx{x}\widetilde\calL(\phi^\pm,Q,P)},
\nonumber\\
\widetilde\calL=&&
\frac{N}{2\cc}
\left[
 \tr(Q^2+P^2)-\omega\tr s(Q+iP)-\omega\tr s(Q-iP)
\right]
+\calL_{\rm p}+\calL_E
-\tr
 \left( 
   Q\sqrts{}\rho\sqrts{}+P\sqrts{}\sigma\sqrts{}
 \right)
\nonumber\\
&&\qquad\qquad
-\frac{N}{\cc}
\left[
 \tr h^+(Q-iP)+\tr h^-(Q+iP)-\omega\tr s(h^++h^-)-2\tr h^+h^-
\right],
\label{TilLag}%------------------------------------------------------
\end{eqnarray}
where we neglect $\omega^2$ term, which vanishes after the replica
limit ($m\rightarrow 0$).
Integrating out the fields $\phi^\pm$, 
we end up with the following action,
%--------------------------------------------------------------------
%   effective action
%--------------------------------------------------------------------
\begin{eqnarray}
&&
\calZ=\int\calD[Q]\calD[P]
 e^{-S_{\rm a}},
\nonumber\\
&&
S_{\rm a}=  \intx{x}\frac{N}{2\cc}
\left[
 \tr(Q^2+P^2)-\omega\tr s(Q+iP)-\omega\tr s(Q-iP)
\right]
+\frac{N}{2}\Tr\Ln C
\nonumber\\
&&\qquad\qquad
-\int d^2x\frac{N}{\cc}
\left[
  \tr h^+(Q-iP)+\tr h^-(Q+iP)
 -\omega\tr s(h^++h^-)-2\tr h^+h^-
\right],
\label{MatFieAct}%---------------------------------------------------
\end{eqnarray}
where $\Tr$ means the trace in $x$-space as well as
$\tr$ and the trace in $\alpha$-space, and $C$ is defined by
%--------------------------------------------------------------------
%   matrix C
%--------------------------------------------------------------------
\begin{equation}
C_{\alpha\alpha'\ak\ak'}=
\left(
 \begin{array}{cc}
  \left[
    E_k\delta_{aa'}\delta_{kk'}-(Q_{aa'}^{kk'}+iP_{aa'}^{kk'})
  \right]\delta_{\alpha\alpha'}
  &-h_{0\alpha\alpha'}\delta_{aa'}\delta_{kk'} 
  \\
  h_{0\alpha\alpha'}\delta_{aa'}\delta_{kk'} 
  &
  \left[
   E_k\delta_{aa'}\delta_{kk'}-(Q_{aa'}^{kk'}-iP_{aa'}^{kk'})
  \right]
  \delta_{\alpha\alpha'}
 \end{array}
\right) ,
\end{equation}
with $E_k=(-)^{q+1}E$. The Green functions are given by
%-----------------------------------------------------------------------
%   Green functions
%-----------------------------------------------------------------------
\begin{eqnarray}
&&
\overline{G}_\pm^k(x)=\lim_{m\rightarrow0}\frac{N}{\cc}
\left[
 \left\langle 
  \left(Q\mp iP \right)_{aa}^{kk}(x)
 \right\rangle -\omega s_k
\right],
\nonumber\\
&&
\overline{K}_{rr'}^{kk'}(x,x')=\lim_{m\rightarrow0}
\left(\frac{N}{2\cc}\right)^2
\left\langle 
  \left(Q-r'iP \right)_{a'a}^{k'k}(x) 
  \left(Q-r iP \right)_{aa'}^{kk'}(x')
\right\rangle
-\frac{N}{2\cc}(1-\delta_{rr'})\delta(x-x').
\end{eqnarray}
\begin{multicols}{2}%--------------------------------------------!!!
%--------------------------------------------------------------------
%                   SADDLE POINT SOLUTION
%--------------------------------------------------------------------
\subsection{Saddle point solution} \label{s:Sad}

Most dominant contribution to the action, 
which is assumed to be independent of the coordinate, 
should satisfy the saddle point equations,
%--------------------------------------------------------------------
%   saddle-point equation
%--------------------------------------------------------------------
\begin{eqnarray}
&&
\frac{1}{\cc}
\left[
\left(Q\pm iP\right)-\omega s
\right]
\nonumber\\&&\quad%----------------------------------------------!!!
=\frac{2\left[E_k-(Q\pm iP)\right]}
{\left[E_k-(Q+iP)\right]\left[E_k-(Q-iP)\right]+\nabla^2}(x,x),
\label{GenSad}%------------------------------------------------------
\end{eqnarray}
where relation $h_0^2=\nabla^2\delta_{\alpha\beta}$ is used, and the 
factor 2 in the numerator comes from the summation 
with respect to $\alpha=1,2$.
If $\omega=0$, it is expected to be of the form
%--------------------------------------------------------------------
%   saddle-point eq. for \omega=0
%--------------------------------------------------------------------
\begin{equation}
P=0,\quad Q=(E_{0k}+s_kq)\delta_{aa'}\delta_{kk'} .
\end{equation}
Then, Eq. (\ref{GenSad}) reduces to one equation,
%--------------------------------------------------------------------
%   one equation
%--------------------------------------------------------------------
\begin{eqnarray}
\frac{1}{\cc}Q=\frac{2(E_k-Q)}{(E_k-Q)^2+\nabla^2}(x,x).
\end{eqnarray}
At zero energy
this equation is easily solved as, setting $E_k=E_{0k}=0$, 
%--------------------------------------------------------------------
%   solution at zero energy
%--------------------------------------------------------------------
\begin{equation}
q=\Lambda(e^{\frac{2\pi}{\cc}}-1)^{-\frac{1}{2}}
\sim\Lambda e^{-\frac{\pi}{\cc}},
\end{equation}
where $\Lambda$ is a cut-off mentioned in Section \ref{s:Mod}.
Therefore, DOS is given by
%--------------------------------------------------------------------
%   DOS at saddle-point level
%--------------------------------------------------------------------
\begin{equation}
\rho(E=0)\sim\frac{2N\Lambda}{\pi\cc} e^{-\frac{\pi}{\cc}}.
\end{equation}
This equation shows that, 
for arbitrary coupling $\cc$, nontrivial solution 
exists, and the GL($4m$,R) symmetry is spontaneously broken.

%--------------------------------------------------------------------
%                      NONLINEAR SIGMA MODEL
%--------------------------------------------------------------------
\section{Derivation of nonlinear sigma model}\label{s:Der}

In this section, we set $E=0$ and concentrate on the band center.
We will divide the fields into longitudinal and Goldstone modes
and derive their transformation law under 
GL($4m$,R) transformation (\ref{GlTra}).
This will turn out to be of importance when we derive a 
gauge-independent effective action of the Goldstone mode
integrating out the longitudinal mode.

%--------------------------------------------------------------------
%                      Transformation properties
%--------------------------------------------------------------------
\subsection{Transformation properties} \label{s:TraPro}

So far we have shown that the Lagrangian
is invariant under the global GL($4m$,R) transformation (\ref{GlTra})
when $\omega=0$.
Hence the action for the auxiliary fields (\ref{MatFieAct}) 
should keep the same invariance.
To study the transformation property of the auxiliary fields, 
let us consider the Lagrangian (\ref{TilLag}). 
Due to the coupling term between $\phi^\pm$ and $Q, P$,
the transformation (\ref{GlTra}) for the quadratic terms,
$
\phi_{\sigma\alpha}^+\phi_{\sigma\alpha}^+
\rightarrow
\calO\phi_{\sigma\alpha}^+\phi_{\sigma\alpha}^+\trns\calO
$
and
$
\phi_{\sigma\alpha}^-\phi_{\sigma\alpha}^-
\rightarrow
s\trns\calO^{-1}s\phi_{\sigma\alpha}^-\phi_{\sigma\alpha}^-s\calO^{-1}s
$
induce the transformation of $Q$ and $P$,
%--------------------------------------------------------------------
%   transformation of Q+-iP
%--------------------------------------------------------------------
\begin{eqnarray}
&&
Q+iP\rightarrow
 \sqrts{-}\trns\calO^{-1}\sqrts{}(Q+iP)
 \sqrts{}\calO^{-1}\sqrts{-},
\nonumber\\
&&
Q-iP\rightarrow
 \sqrts{}\calO\sqrts{-}(Q-iP)
 \sqrts{-}\trns\calO\sqrts{}.
\label{QPTra}%--------------------------------------------------------
\end{eqnarray} 
In the same way, the source terms should transform as
%--------------------------------------------------------------------
%   transformation of h
%--------------------------------------------------------------------
\begin{eqnarray}
&&
h^+\rightarrow
 \sqrts{-}\trns\calO^{-1}\sqrts{} h^+
 \sqrts{}\calO^{-1}\sqrts{-},
\nonumber\\
&&
h^-\rightarrow
 \sqrts{}\calO\sqrts{-}h^-
 \sqrts{-}\trns\calO\sqrts{}.
\end{eqnarray}
Note that the saddle point solution is invariant under the 
transformation $\calo\in{\rm O}(4m)$. Therefore, we can decompose
the $Q$ and $P$ fields as follows
%--------------------------------------------------------------------
%   poler-decomposition
%--------------------------------------------------------------------
\begin{eqnarray}
&&
Q+iP=
 \sqrts{-}\trns T^{-1}\sqrts{}\Rm^+\sqrts{} T^{-1}\sqrts{-},
\nonumber\\
&&
Q-iP=
 \sqrts{} T\sqrts{-}\Rm^-\sqrts{-}\trns T\sqrts{},
\label{PolDec}%------------------------------------------------------
\end{eqnarray} 
where the field $T(x)\in{\rm GL}(4m,{\rm R})/{\rm O}(4m)$
describe the Goldstone mode, 
and longitudinal fields are parameterized as
%--------------------------------------------------------------------
%   Longitudinal modes
%--------------------------------------------------------------------
\begin{eqnarray}
&&
\Rm^\pm(x)=\Rm_0^\pm+sq,
 \quad
\Rm_0^\pm=\Qm(x)\pm i\Pm(x),
\nonumber\\
&&
Q_L=\frac{1}{2}(R-sRs), \quad P_L=\frac{1}{2}(R+sRs),
%\trns\Qm=\Qm=-s\Qm s,\quad \trns\Pm=\Pm=s\Pm s.
\label{LonMod}%-------------------------------------------------------
\end{eqnarray} 
with a real and symmetric matrix $R$.
The constant imaginary shift $sq$ as well as the last two 
parameterizations ensure                    \cite{foot1}%---------!!!
%\footnote{ %------------------------------------------------------!!!
%} %--------------------------------------------------------------!!!
%------------------------------------------------------------------- 
that the integration over $\phi^\pm$ fields 
converges in Eq. (\ref{MatFieAct}).
Moreover, it is easily shown that 
the degrees of freedom are equivalent: 
The two real and symmetric matrices $Q$ and $P$ are now 
converted to the two real and symmetric matrices $T$ 
($\trns T=T$ is one of possible gauges) and $R$.
The following identities are hold;
\begin{equation}
sL_0^\pm s=-L_0^\mp.
\label{Ide}%---------------------------------------------------------
\end{equation}

Next, examine the transformation of $T$ and $L$ fields.
First, we have to consider the action of 
$\calO\in$GL($4m$,R) on $T$, 
%--------------------------------------------------------------------
%   product of GT
%--------------------------------------------------------------------
\begin{equation}
\calO T(x)=T'(x)\calo(T(x),\calO),
\label{ProT}%---------------------------------------------------------
\end{equation}
where 
$T'(x)\in{\rm GL}(4m,{\rm R})/{\rm O}(4m)$ 
and 
$\calo(T,\calO)\in{\rm O}(4m)$.
It should be noted that the field $\calo(T,\calO)$ 
is a nonlinear function of $T(x)$. 
Applying the transformation (\ref{QPTra}) to
(\ref{LonMod}) and considering Eq. (\ref{ProT}),
we have the following transformation laws
%--------------------------------------------------------------------
%   transformation of T and L
%--------------------------------------------------------------------
\begin{eqnarray}
&&
T\rightarrow\calO T\calo^{-1}(T,\calO),
\nonumber\\
&&
\Rm^+\rightarrow\sqrts{-}\calo(T,\calO)\sqrts{}
\Rm^+\sqrts{}\calo^{-1}(T,\calO)\sqrts{-},
\nonumber\\
&&
\Rm^-\rightarrow\sqrts{}\calo(T,\calO)\sqrts{-}
\Rm^-\sqrts{-}\calo^{-1}(T,\calO)\sqrts{}.
\end{eqnarray}
\end{multicols}\noindent%----------------------------------------!!!
Under the change of variables (\ref{PolDec}), 
the action (\ref{TilLag}) yields,
%--------------------------------------------------------------------
%   effective action
%--------------------------------------------------------------------
\begin{eqnarray}
&&
\calZ=\int\calD[T]\calD[L]I[L]
 e^{-S_{\rm a}},
\nonumber\\
&&
S_{\rm a}=\intx{x}\frac{N}{2\cc}
\left[
 \tr\Rm^+\Rm^-
 -\omega\tr(\trns TT)^{-1}\sqrts{}\Rm^+\sqrts{}
 +\omega\tr(\trns TT)\sqrts{-}\Rm^-\sqrts{-}
\right]
+\frac{N}{2}\Tr\Ln C
\nonumber\\
&&\qquad
-\intx{x}\frac{N}{\cc}
\Big[
  \tr h^+\sqrts{}T\sqrts{-}\Rm^-\sqrts{-}\trns T\sqrts{}
 +\tr h^-\sqrts{-}\trns T^{-1}\sqrts{}\Rm^+\sqrts{}T^{-1}\sqrts{-}
%\nonumber\\&&\qquad\qquad\qquad\qquad%-----------------------------!!!
 -\omega\tr s(h^++h^-)-2\tr h^+h^-
\Big],
\end{eqnarray}
where $I[L]$ is a measure and $C$ is converted into,
after the gauge-transformation
%--------------------------------------------------------------------
%   matrix C
%--------------------------------------------------------------------
\begin{equation}
C=
\left(
\begin{array}{cc}
 -\Rm^+
 &
 -\sqrts{-}\widetilde\gamma_\mu(\partial_\mu+v_\mu-a_\mu)\sqrts{}
 \\
 \sqrts{}\widetilde\gamma_\mu(\partial_\mu+v_\mu+a_\mu)\sqrts{-}
 &
 -\Rm^-
\end{array}
\right),
\end{equation}
with
%--------------------------------------------------------------------
%   vector-like potential
%--------------------------------------------------------------------
\begin{equation}
\left.
 \begin{array}{cc}
  v_\mu\\ a_\mu
 \end{array}
\right\}
=\frac{1}{2}
\left(
 T^{-1}\partial_\mu T\pm \trns T\partial_\mu\trns T^{-1}
\right) .
\end{equation}
It is easily verified that the action is invariant
under the local O($4m$) transformation as well as the global
GL($4m$,R) transformation\cite{BKY}, 
except for the symmetry-breaking terms.
The transformation laws of $v$ and $a$ fields are 
%--------------------------------------------------------------------
%   transformation of vector-like potential
%--------------------------------------------------------------------
\begin{eqnarray}
&&
v_\mu\rightarrow \calo v_\mu \calo^{-1}+\calo\partial_\mu\calo^{-1},
%\nonumber\\
%&&
\quad
a_\mu\rightarrow\calo a_\mu\calo^{-1},
\end{eqnarray}
which means that $v$ is a gauge field associated with 
the hidden local O($4m$) symmetry\cite{BKY}.
Actually, the hidden local O($4m$) symmetry reflects the fact 
that the parameterization of the symmetric space GL($4m$,R)/O($4m$)
is not unique. 
Another comment is that the breaking terms are not $g$-invariant,
but are gauge-invariant.
Therefore, the total Lagrangian including the breaking terms is 
gauge-invariant, as it should be.
%--------------------------------------------------------------------
%                         Derivative expansion
%--------------------------------------------------------------------
\subsection{Derivative expansion} \label{s:DerExp}

Integration over the massive mode yields an effective action for 
the Goldstone mode.
To this end, let us divide the action into two parts,
%--------------------------------------------------------------------
%   divide the action into two parts
%--------------------------------------------------------------------
\begin{eqnarray}
&&
\calZ=\int\calD[T]\calD[L]
 e^{-S_L(L)-\delta S(T,L)},
\nonumber\\
&&
S_L(L)=\intx{x}\frac{N}{2\cc}\tr\Rm^+\Rm^-
+\frac{N}{2}\Tr\Ln A
-\Ln I[L] ,
\label{LonLag}\\ %---------------------------------------------------
&&
\delta S(T,L)=\frac{N}{2}\Tr\Ln(1+A^{-1}B)
+\intx{x}\frac{N}{2\cc}
\left[
 -\omega\tr(\trns TT)^{-1}\sqrts{}\Rm^+\sqrts{}
 +\omega\tr(\trns TT)\sqrts{-}\Rm^-\sqrts{-}
\right] 
\nonumber\\
&&\qquad
-\intx{x}\frac{N}{\cc}
\Big[
  \tr h^+\sqrts{}T\sqrts{-}\Rm^-\sqrts{-}\trns T\sqrts{}
 +\tr h^-\sqrts{-}\trns T^{-1}\sqrts{}\Rm^+\sqrts{}T^{-1}\sqrts{-}
%\nonumber\\&&\qquad\qquad\qquad\qquad%-----------------------------!!!
 -\omega\tr s(h^++h^-)-2\tr h^+h^-
\Big] ,
\label{TraLag}%---------------------------------------------------
\end{eqnarray}
where
%--------------------------------------------------------------------
%   A and B matrices
%--------------------------------------------------------------------
\begin{eqnarray}
%&&
A=
\left(
\begin{array}{cc}
 -\Rm^+
 &
 -\widetilde\gamma_\mu\partial_\mu\
 \\
 \widetilde\gamma_\mu\partial_\mu
 &
 -\Rm^-
\end{array}
\right),\quad
%\nonumber\\
%&&
\quad
B=
\left(
\begin{array}{cc}
 0
 &
 -\sqrts{-}\widetilde\gamma_\mu(v_\mu-a_\mu)\sqrts{}
 \\
 \sqrts{}\widetilde\gamma_\mu(v_\mu+a_\mu)\sqrts{-}
 &
 0
\end{array}
\right) .
\end{eqnarray}
It is reasonable to
define effective Lagrangian for $T$ field as 
$S_T\equiv\langle\delta S(T,L)\rangle$, where $\langle\cdots\rangle$
means the expectation value with respect to $S_L$.
Since we are now interested in long distance behavior of the 
Goldstone mode, we will expand these nonlocal Lagrangian up to 
quadratic order in fields, and will
derive the effective NLSM for the Goldstone mode
up to O($N^0$) for large $N$ system.
%--------------------------------------------------------------------
%                   On the longitudinal mode
%--------------------------------------------------------------------
\begin{multicols}{2}%--------------------------------------------!!!
\subsubsection{Longitudinal mode} \label{s:LonMod}

In this subsection, we derive the leading Lagrangian 
of the longitudinal mode.
In order to get an O($N^0$) NLSM of the Goldstone mode, 
it is enough to derive the Lagrangian
of the longitudinal mode up to O($N$).
Hence, the measure term in Eq. (\ref{LonLag}) can be neglected.
The first term in Eq. (\ref{LonLag}) is expanded as
\begin{equation}
\tr\Rm^+\Rm^-=\tr L_0^+L_0^-+q\tr s(L_0^++L_0^-)-4mq^2,
\label{LonLagFir}%----------------------------------------------------
\end{equation} 
where the last term can be neglected, 
since it vanishes after the replica limit.
The second term, $\Tr\Ln A$, can be expanded as
%--------------------------------------------------------------------
%   expansion of Tr Ln A
%--------------------------------------------------------------------
\begin{eqnarray}
&&
\Tr\Ln A
=\Tr\Ln A_0+\Tr\Ln (1-A_0^{-1}L_0)
\nonumber\\
&&
\sim\Tr\Ln A_0-\Tr A_0^{-1}L_0-\frac{1}{2}\Tr A_0^{-1}L_0A_0^{-1}L_0,
\label{TrLn1}%-------------------------------------------------------
\end{eqnarray}
where
%--------------------------------------------------------------------
%   A, A_0, L_0
%--------------------------------------------------------------------
\begin{eqnarray}
&&
A=A_0-L_0,
\nonumber\\
&&
A_0=
\left(
\begin{array}{cc}
 -sq & -\widetilde\gamma_\mu\partial_\mu \\
 \widetilde\gamma_\mu\partial_\mu &-sq
\end{array}
\right) ,
\quad
L_0=
\left(
\begin{array}{cc}
 \Rm_0^+&0\\0&\Rm_0^-
\end{array}
\right) .
\end{eqnarray}
The linear term with respect to $L_0$ in Eq. (\ref{TrLn1}) 
cancel out that in Eq. (\ref{LonLagFir}), as expected. 
Collecting the quadratic terms 
in Eqs. (\ref{LonLagFir}) and (\ref{TrLn1}), 
we finally have
\begin{equation}
S_L(L)\sim\frac{1}{4d}\intx{x}
\left(
\tr\partial_\mu L_0^+\partial_\mu L_0^-+h_L\tr L_0^+L_0^-
\right) ,
\label{LonAct}%------------------------------------------------------
\end{equation}
where
\begin{equation}
\frac{1}{d}=\frac{N}{6\pi q^2},\quad h_L=12q^2.
\end{equation}
The derivation of this action
is  outlined in Appendix \ref{a:Cal}.
%--------------------------------------------------------------------
%                       Goldstone mode
%--------------------------------------------------------------------
\subsubsection{Goldstone mode} \label{s:TraMod}

In order to derive the NLSM for the Goldstone mode,
let us expand the action $\delta S$ in Eq. (\ref{TraLag})
up to second order with respect to the derivatives.
In this leading order, $v$ does not appear
because it is not gauge covariant: It enters the fourth order 
in the form of covariant derivative $\partial_\mu-v_\mu$.
Thus, in the second order, we can simply neglect $v$.
%----------------------------------------------------------------------
%                         Figure: diagram 1
%----------------------------------------------------------------------
\begin{figure}[htb] %[e]
\vspace{5mm}
\centerline{
\epsfxsize=30mm\epsfbox{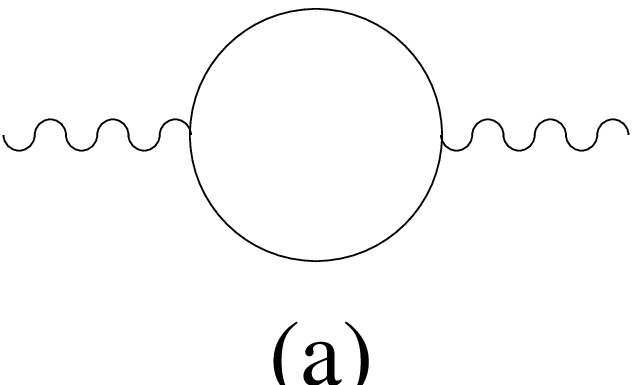}
           }
%\quad
\vspace{5mm}
\centerline{
\epsfxsize=30mm\epsfbox{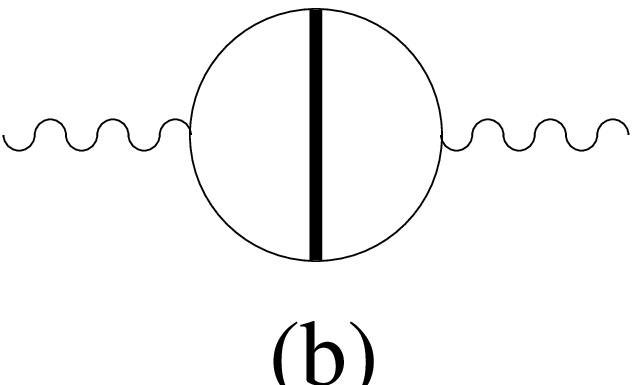}
           }
%\quad
\vspace{5mm}
\centerline{
\epsfxsize=30mm\epsfbox{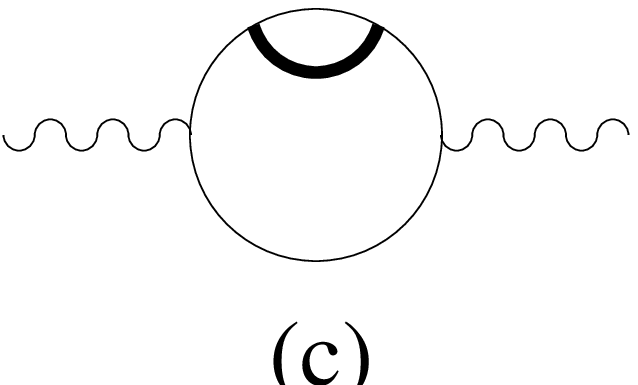}
           }
\vspace{5mm}
\caption{
Diagrams contributing (a) to order $N$ and  
(b), (c) to order $N^0$. 
Wavy line denotes $a_\mu$, while
thin and thick lines denote, respectively, the propagator
of $A_0$ and of $L$ field, summarized in Appendix \ref{a:Cal}.
        }
\label{f:Dia}%--------------------------------------------------------
\end{figure}\noindent
Using 
%----------------------------------------------------------------------
%   Expansion of Tr Ln
%----------------------------------------------------------------------
\begin{equation}
\Tr\Ln(1+A^{-1}B)\sim\Tr A^{-1}B-\frac{1}{2}\Tr A^{-1}BA^{-1}B,
\label{TrLn2}%---------------------------------------------------------
\end{equation}
as well as
%----------------------------------------------------------------------
%   Expansion of A
%----------------------------------------------------------------------
\begin{equation}
A^{-1}\sim A_0^{-1}+A_0^{-1}L_0A_0^{-1}+A_0^{-1}L_0A_0^{-1}L_0A_0^{-1},
\label{ExpA}%----------------------------------------------------------
\end{equation} 
\end{multicols}\noindent%------------------------------------------!!!
it turns out that diagrams summarized in Fig. \ref{f:Dia}
contribute to O($N^0$).
Accordingly, we reach 
%----------------------------------------------------------------------
%   Action of the Goldstone mode
%----------------------------------------------------------------------
\begin{eqnarray}
S_T&&
=\intx{x}
\left\{
 \frac{1}{b}\tr a_\mu^2-\frac{1}{c}\tr^2a_\mu
+\frac{N}{2\cc}\omega q\tr
\left[
 (\trns TT)^{-1}+(\trns TT)
\right]
\right\}
\nonumber\\
&&\qquad
-\intx{x}\frac{N}{\cc}
\left[
  q\tr h^+\sqrts{}(T\trns T)\sqrts{}
 -q\tr h^-\sqrts{-}(T\trns T)^{-1}\sqrts{-}
 -\omega\tr s(h^++h^-)
 -2\tr h^+h^-
\right],
\label{TraModAct}%-----------------------------------------------------
\end{eqnarray}
where coupling constants are given by
%----------------------------------------------------------------------
%   coupling constants
%----------------------------------------------------------------------
\begin{eqnarray}
&&
\frac{1}{b}=\frac{N}{4\pi}
%\left
% (1-e^{-\frac{2\pi}{\cc}}
%\right)
-\frac{d_{\rm b}-d_{\rm c}(1+4m)}{4\pi}
+\mbox{O}(N^{-1}),
\nonumber\\
&&
\frac{1}{c}=
\frac{d_{\rm b}}{4\pi}
+\mbox{O}(N^{-1}),
\nonumber\\
&&
d_{\rm b}\sim 0.4593, \quad d_{\rm c}\sim 0.1453.
\label{IniCouCon}%-----------------------------------------------------
\end{eqnarray}
The contribution from the diagrams (b) and (c) is denoted by
$d_{\rm b}$ and $d_{\rm c}$, respectively.
An outline of the calculation is summarized in Appendix \ref{a:Cal}.
Some comments may be in order.
Firstly, 
what is characteristic in this Lagrangian is that there appears
$\tr^2$-term as well as the principal term.
As we shall see later, this term plays an important role in the 
scaling property of DOS.
Secondly,
at the leading order of the large $N$ expansion, the coupling 
constant $1/c$ is 0, and in the next leading order,
it is a finite positive values. 
This sign is quite relevant to the renormalization group flow,
as we shall see momentarily.
Lastly,
this Lagrangian is manifestly gauge-invariant.
If we fix the gauge as $\trns T=T$, the action becomes
%----------------------------------------------------------------------
%   action of Goldstone mode after gauge-fixing
%----------------------------------------------------------------------
\begin{eqnarray}
&&
S_T=\intx{x}
\left[
 -\frac{1}{4b}\tr\partial_\mu T^{-2}\partial_\mu T^2
 -\frac{1}{4c}\tr^2T^{-2}\partial_\mu T^2
 +\frac{N}{2\cc}\omega q\tr
\left(
  T^{-2}+T^2
\right)
\right]
\nonumber\\
&&\hspace{20mm}
-\intx{x}\frac{N}{\cc}
\left[
  q\tr h^+\sqrts{}T^2\sqrts{}
 -q\tr h^-\sqrts{-}T^{-2}\sqrts{-}
 -\omega\tr s(h^++h^-)
 -2\tr h^+h^-
\right] .
\end{eqnarray}
This is an action of GL($4m$,R)/O($4m$) NLSM,
similar to the one for the two-sublattice model 
with broken time-reversal symmetry derived by Gade.
The Green function is
\begin{eqnarray}
&&
\overline{G}_\pm^k(x)=\lim_{m\rightarrow0}\frac{N}{\cc}qs_k
\left[
 \langle (T^{\pm2})_{aa}^{kk}\rangle -\omega
\right],
\nonumber\\
&&
\overline{K}_{rr'}^{kk'}(x,x')=\lim_{m\rightarrow0}
\left(\frac{N}{2\cc}\right)^2
\left\langle 
  \left(\sqrts{}T^2\sqrts{}\right)^{r'}\,\!_{a'a}^{k'k}(x) 
  \left(\sqrts{}T^2\sqrts{}\right)^{r }\,\!_{aa'}^{kk'}(x')
\right\rangle
-\frac{N}{2\cc}(1-\delta_{rr'})\delta(x-x').
\end{eqnarray}

%--------------------------------------------------------------------
%                         RENORMALIZATION GROUP
%--------------------------------------------------------------------
\section{Renormalization Group} \label{s:Ren}

By the use of the relation\cite{Hik} between
the noncompact symmetric space GL($4m$,R)/O($4m$)
and the compact symmetric space U($4m$)/SO($4m$), 
it is easy to write down the beta functions of the 
renormalization group\cite{GadWeg,Weg}
%--------------------------------------------------------------------
%   beta functions
%--------------------------------------------------------------------
\begin{eqnarray}
\beta_b(b,c)=
&&
%\textstyle{
% \varepsilon b
\frac{1}{2} n b^2
-\frac{1}{4}\left(\frac{1}{2}n^2+n\right) b^3
+\frac{1}{8}\left(\frac{3}{8}n^3+\frac{5}{4}n^2+n\right) b^4
%          }
\nonumber\\
&&
%\textstyle{
-\frac{1}{16}
\left[
  \frac{19}{48}n^4
 +\left(\frac{43}{24}-\frac{3}{16}\zeta(3)\right)n^3
 +\left(\frac{9}{4}+\frac{3}{8}\zeta(3)\right)n^2
 +\frac{1}{2}n
\right]b^5-\mbox{O}(b^6),
%         }
\nonumber\\
\beta_c(b,c)=
&&
%\varepsilon c
 c^2\frac{\beta_b(b,c)}{nb^2},
\end{eqnarray}
\begin{multicols}{2}\noindent%-----------------------------------!!!
where $n=4m$ and $\zeta$ is the Riemann zeta function. 
Here we changed the sign of $b$ of beta functions 
in Ref. \cite{GadWeg}, 
because the present symmetric
space is noncompact.
We also took the normalization of $b$ and $c$ into account
by calculating the renormalization constants explicitly 
up to two-loop order.
The scaling equations are, therefore, given by,
in the replica limit ($m\rightarrow0$), 
%--------------------------------------------------------------------
%   scaling equations
%--------------------------------------------------------------------
\begin{eqnarray}
&&
\diff{b}{l}=0,
\nonumber\\
&&
\diff{c}{l}=
%\textstyle{
-\frac{1}{2}c^2
\left[   
 1-\frac{b}{2}+\left(\frac{b}{2}\right)^2
  -\frac{1}{2}\left(\frac{b}{2}\right)^3 +\mbox{O}(b^4)
\right],
%          }
\label{ScaEq}%--------------------------------------------------------
\end{eqnarray}
where $l=\ln L$ with $L$ being the length of the system and 
coupling constants are rescaled as 
$b,c\rightarrow\frac{b}{2\pi},\frac{c}{2\pi}$,
whose initial values are given by
%--------------------------------------------------------------------
%   initial values
%--------------------------------------------------------------------
\begin{eqnarray}
&&
b_0=\frac{2}{N-(d_{\rm b}-d_{\rm c})}\sim\frac{2}{N-0.314},
\nonumber\\
&&
c_0=\frac{2}{d_{\rm b}}\sim4 .
\label{ResIni}%-------------------------------------------------------
\end{eqnarray}

Let us now discuss the flow of the coupling constants.
First of all, vanishing beta function for $b$ implies
that $E=0$ state is delocalized\cite{Gad}. For $E\ne0$, 
the universality
class O($2m,2m$)/O($2m)\times$O($2m$) suggests that such states
are localized, as discussed in Section \ref{s:Rep}. 
Therefore, only the band center is 
just on the critical point for the present model.
Actually, this is in good agreement with 
the numerical calculation of HWKM for the $N=1$ model.

To clarify the flow of $c$,
let us first consider the case where $N$ is large enough to ignore
the O($N^{-1}$) in Eq. (\ref{IniCouCon}).
In this case, $b$ is quite small while $c$ is positive.
The scaling equation (\ref{ScaEq}) then
tells us that $c$ scales to 0.
Therefore, large $N$ systems behave similar to the class 
with broken time reversal symmetry\cite{Gad}.
Namely, since the $\zeta$ function is given by \cite{GadWeg}
%--------------------------------------------------------------------
%   zeta function
%--------------------------------------------------------------------
\begin{eqnarray}
\zeta&&=\frac{n+1}{2}b+\frac{b^2}{c-nb}
 -\frac{3}{64}
 \left( 4-3n^2-n^3\right)b^3
\nonumber\\
&&\rightarrow 
\frac{b}{2}+\frac{b^2}{c}-\frac{3}{16}b^3\quad(m\rightarrow0),
\end{eqnarray}
the DOS at the zero energy 
diverges under the change of the length scale;
%--------------------------------------------------------------------
%   divergent DOS
%--------------------------------------------------------------------
\begin{equation}
\rho\propto q
 \exp\left(\int^\infty\!\!\zeta dl\right)
\rightarrow\infty.
\end{equation}
On the one hand,
this may be likely, since Oppermann and Wegner already derived 
a singular behavior of the Green function 
of the two-sublattice model such as $1/N(d-2)$,
using the large $N$ expansion.
On the other hand, 
the present divergence of the DOS suggests a discontinuity 
with respect to $\cc$ at $\cc=0$, 
since DOS is exactly zero at the band center when $\cc=0$.

However, the numerical calculation for the $N=1$ system
by HWKM suggests
the convergence of the DOS even for rather large $\cc$. 
Therefore, small $N$ systems, at least $N=1$ system, 
should belong to a different fixed point.
Actually in Eq. (\ref{ScaEq}), there is 
a nontrivial zero for $\beta_c$ at $b=b_{\rm c}\sim3.087$.
Accordingly,
if $b_0<b_{\rm c}$, $c$ flows to 0, while if  $b_0>b_{\rm c}$, 
$c$ flows to infinity, provided that $c_0>0$.
The scaling equation (\ref{ScaEq}) up to four loop order shows, 
therefore,  
the existence of a strong-coupling fixed point in addition to the 
weak-coupling fixed point $c=0$ mentioned above.
Since $c$ diverges at a certain length scale $l=l_{\rm c}$, 
the DOS in the strong-coupling limit may scale as
%--------------------------------------------------------------------
%   DOS in the strong-coupling limit
%--------------------------------------------------------------------
\begin{equation}
\rho\propto q
 \exp\left(\int^{l_{\rm c}}\!\!\zeta dl\right)
\propto q.
\end{equation}
Namely, DOS is expected to be convergent in the strong-coupling phase.
The initial value of the present calculation is
$b_0=2.92$ for $N=1$, which
is still in the weak-coupling regime, to be sure, 
but lies quite near to the zero of the beta function.

Of course, precise studies of this strong-coupling regime
are beyond our scope, since we are based on the perturbations:
The initial values $b_0$ in Eq. (\ref{ResIni})
is valid only for large $N$ cases.
The zero of the beta function depends, moreover, on the order
of the loop expansion. Namely,
$b_{\rm c}$ is computed as
$b_{\rm c}=2$ and  3.087 
for two- and four-loop order, respectively,
but no $b_{\rm c}$ for one- and three-loop order.
However, numerical calculations by HWKM for the $N=1$ system
suggests the convergence of the DOS, 
which tells that such model cannot be in the weak-coupling 
fixed point.

Therefore, we conjecture the existence of a critical $N$ which
separate the strong- and the weak-coupling phases.
It is quite interesting to observe the weak-coupling one
which is suggested in this paper for the first time,
and to determine the phase diagram more precisely in 
the $\cc$-$N$ plane by nonperturbative methods or
by numerical calculations.

%--------------------------------------------------------------------
%                       SUMMARY
%--------------------------------------------------------------------
\section{Summary} \label{s:Sum}

We have investigated the 2D random hopping fermion model 
proposed by HWKM,
using field theoretical treatments developed by Gade.
Starting from the tight-binding Hamiltonian,
we have firstly derived the continuum Dirac Hamiltonian which 
effectively describes the band center of the lattice model,
and next constructed the generating functional of the
Green functions averaged-over the disorder.
It has been shown that the symmetry group is GL($4m$,R), which is
spontaneously broken to O($4m$).
Integrating out the massive modes, we have derived a nonlinear
sigma model on the symmetric space GL($4m$,R)/O($4m$), describing 
the Goldstone mode.
The beta functions of the renormalization group show that 
the band center is a random critical point, where 
the density of state diverges for large $N$ systems.
This corresponds to a weak-coupling fixed point, 
which may share basic properties with the two sublattice model
studied by Oppermann and Wegner.

However, due to a nontrivial zero of the beta function,
it is likely that the small $N$ system is in the 
strong-coupling limit.
This fixed point is still critical 
but presumably with convergent density of state.
This fixed point may occur at a large value of the coupling 
constant, so that it is beyond our perturbative theory.
It is, therefore, quite interesting to study
the phase diagram more precisely 
by nonperturbative methods or by numerical calculations.

%--------------------------------------------------------------------
%                       ACKNOWLEDGEMENTS
%--------------------------------------------------------------------
\acknowledgements

The author would like to thank Y. Hatsugai, Y. Morita,
A. Altland, N. Kawakami, and especially M. R. Zirnbauer 
for fruitful discussions and a lot of comments.
He also would like to thank J. Zittartz for helpful comments 
as well as his kind hospitality.
The author is supported by JSPS Postdoctoral Fellowships
for Research Abroad.

\end{multicols}%--------------------------------------------------!!!

%--------------------------------------------------------------------
%                       APPENDIX
%--------------------------------------------------------------------
\appendix
\section{Calculations of the leading order actions for 
the longitudinal and the Goldstone modes}
\label{a:Cal}%-------------------------------------------------------

The basic propagator $A_0^{-1}$ is
%--------------------------------------------------------------------
%   propagator A_0
%--------------------------------------------------------------------
\begin{equation}
A_{0\,\alpha\alpha'\ak\ak'}^{-1}(x,x')=
\delta_{aa'}\delta_{kk'}
\left(
 \begin{array}{cc}
  \delta_{\alpha\alpha'}s_k\Ad(x,x')
 &-\Aod{\alpha\alpha'}(x,x')
 \\
  \Aod{\alpha\alpha'}(x,x')
 &\delta_{\alpha\alpha'}s_k\Ad(x,x')
 \end{array}
\right) ,
\end{equation}
where,
%--------------------------------------------------------------------
%   elements of propagator of A_0
%--------------------------------------------------------------------
\begin{eqnarray}
&&
\Ad(x,x')=
 \intp{p}\frac{q}{p^2+q^2}e^{ip\cdot(x-x')},
\\
&&
\Aod{\alpha\alpha'}(x,x')=\intp{p}\frac{i}{p^2+q^2}
\left(
\begin{array}{rr}
 p_x & p_y \\ p_y & -p_x
\end{array}
\right)_{\alpha\alpha'}
e^{ip\cdot(x-x')} .
\end{eqnarray}

Let us first consider the longitudinal mode 
in Section \ref{s:LonMod}. 
Main calculation lies in the 
quadratic term with respect to $L_0$ in Eq. (\ref{TrLn1}), 
calculated as follows:
%--------------------------------------------------------------------
%   Tr ALAL
%--------------------------------------------------------------------
\begin{eqnarray}
&&
\Tr A_0^{-1}L_0A_0^{-1}L_0
\nonumber\\
&&
=\intx{x}\intx{x'}
\left[
 2\Ad(x,x')\Ad(x',x)\tr sL_0^+(x')sL_0^+(x)
 -\Aod{\alpha\alpha'}(x,x')\Aod{\alpha'\alpha}(x',x)
 \tr L_0^-(x')L_0^+(x)
\right]
%\nonumber\\&&\qquad\qquad\qquad\qquad%-----------------------------!!!
+(+\rightarrow -)
\nonumber\\
&&
\sim\intx{x'}\intp{p}\intp{p'}
 \frac{2(p\cdot p'-q^2) e^{-i(p-p')\cdot x'}}
      {(p^2+q^2)(p'^2+q^2)}
\intx{x}
\left[
  2\tr L_0^+(x)L_0^-(x)
 -x'^2_\mu\tr\partial_\mu L_0^+(x)\partial_\mu L_0^-(x)
\right],
\end{eqnarray}
where we used Eq. (\ref{Ide}) as well as a local field approximation
$L_0^\pm(x')\sim L_0^\pm(x)+(x'-x)_\mu\partial_\mu L_0^\pm(x)+
\frac{1}{2}(x'-x)_\mu(x'-x)_\nu\partial_\mu\partial_\nu L_0^\pm(x)$.
The first $\tr$ term in the above equation as well as the first one in 
Eq. (\ref{LonLagFir}) give the mass term in Eq. (\ref{LonAct}), 
while the second $\tr$ term becomes the kinetic term.

In order to calculate the diagrams in Fig. \ref{f:Dia}
for the Goldstone mode in Section \ref{s:TraMod},
we need the propagator of the longitudinal mode, given by
%--------------------------------------------------------------------
%   propagator of longitudinal mode
%--------------------------------------------------------------------
\begin{eqnarray}
&&
\left\langle 
 L_{0~\;aa'}^{\pm~kk'}(p)L_{0~\,bb'}^{\mp~ll'}(-p)
\right\rangle
=
\left( 
 \delta_{ab}^{kl}\delta_{a'b'}^{k'l'}
  +\delta_{ab'}^{kl'}\delta_{a'b}^{k'l} 
\right)
 \frac{d}{p^2+h_L},
\nonumber\\
&&
\left\langle 
 L_{0~\;aa'}^{\pm~kk'}(p)L_{0~\,bb'}^{\pm~ll'}(-p)
\right\rangle
=-s^ks^{k'}
\left( 
 \delta_{ab}^{kl}\delta_{a'b'}^{k'l'}
  +\delta_{ab'}^{kl'}\delta_{a'b}^{k'l} 
\right)
 \frac{d}{p^2+h_L}.
\end{eqnarray}
The last equation follows from Eq. (\ref{Ide}).
By using these formulas, we have the following expressions
contributing to the action (\ref{TraModAct})
from each diagram in Fig. \ref{f:Dia},
\begin{eqnarray}
&&
\mbox{(a)}=
N\intp{p}
  \frac{q^2}
       {(p^2+q^2)^2}
 \intx{x}\tr a_\mu^2,
\nonumber\\
&&
\mbox{(b)}=
-Nd\intp{p}\intp{p'}
  \frac{(p\cdot p')^2-2q^2p\cdot p'+q^4}
       {(p^2+q^2)^2(p'^2+q^2)^2[(p-p')^2+h_L]}
 \intx{x}\left(\tr a_\mu^2+\tr^2a_\mu\right),
\nonumber\\
&&
\mbox{(c)}=
2Nd(1+4m)\intp{p}\intp{p'}
  \frac{q^2(2p\cdot p'+p^2)-q^4}
       {(p^2+q^2)^3(p'^2+q^2)[(p-p')^2+h_L]}
 \intx{x}\tr a_\mu^2.
\end{eqnarray}
Coefficients of these terms (integrals over $p$) 
are in principle dependent on $\cc$. 
To illustrate this, let us consider
the first one, for example: 
Since it is convergent, we can integrate it
over the whole 2D momentum space, giving the value $N/4\pi$.
However, for the present model, the cut-off is a physical parameter,
denoting the scale under which the linearization procedure of
the dispersion relation is valid.
Accordingly,
if we introduce the cut-off to the (convergent) integral, 
it turns out to be dependent on $\cc$ through
the saddle point solution $q$ such as 
$\frac{N}{4\pi}\left(1-\frac{q^2}{\Lambda^2+q^2}\right)
=\frac{N}{4\pi}(1-e^{-\frac{2\pi}{g}})$.
However,
for quite small $\cc$, the exponential in this expression
can be neglected again.

\begin{multicols}{2}%---------------------------------------------!!!
%--------------------------------------------------------------------
%   references
%--------------------------------------------------------------------

\end{multicols}%------------------------------------------------!!!

%----------------------------------------------------------------------
%                         Figure: diagram 1
%----------------------------------------------------------------------


\begin{references}
\bibitem{foot1}%--------------------------------------------------!!!
Consider only $\phi^+$ sector, because 
similar arguments hold in the other sector.
Coupling term yields
$
-\trns\phi^+\sqrts{}(Q+iP)\sqrts{}\phi^+
=
\trns\phi^+\trns T^{-1}
\left[q-\sqrts{}(Q_L+iP_L)\sqrts{}\right]T^{-1}\phi^+
$.
In this equation, $q$ basically ensures the convergence 
of the integration over $\phi^+$. For the integration to be really
convergent, $\sqrts{}(Q_L+iP_L)\sqrts{}$ needs being purely
imaginary. Note that 
$
\sqrts{}\left(\begin{array}{rr}x&y\\y&z\end{array}\right)\sqrts{}
=\left(\begin{array}{rr}ix&y\\y&-iz\end{array}\right)
$, 
where $x$, $y$ and $z$ are $2m\times2m$ matrices with the same $p$ 
sector.
Therefore, $Q$ and $P$ should have only diagonal and only off-diagonal
matrix elements in $k$ sector, respectively.
Using
$
s\left(\begin{array}{rr}x&y\\y&z\end{array}\right)s
=\left(\begin{array}{rr}-x&y\\y&-z\end{array}\right)
$,
they can be written as Eq. (\ref{LonMod}).
\end{references}
\end{document}